\documentclass[aps,prl,preprint,tightenlines,superscriptaddress,showpacs,byrevtex]{revtex4}
\usepackage{graphicx} 
\usepackage{dcolumn}  

\newcommand{\aonepibr}{14.9}
\newcommand{\aonepibrstat}{1.6}
\newcommand{\aonepibrsyst}{2.3}
\newcommand{\aonepi}{a_1^{\pm}(1260)\pi^{\mp}}
\newcommand{\aone}{a_1^{\pm}(1260)}
\newcommand{\atwo}{a_2^{\pm}(1320)}
\newcommand{\atwopi}{a_2^{\pm}(1320)\pi^{\mp}}
\newcommand{\aonedecay}{a_1^{\pm}(1260)\to\pi^{\pm}\pi^{\pm}\pi^{\mp}}

\newcommand{\de}{\Delta E}
\newcommand{\mbc}{M_{\rm bc}}

\graphicspath{{ps}}

\begin{document}


\preprint{\vbox{ \hbox{   }
                 \hbox{BELLE-CONF-0728}
}}

\title{ \quad\\[0.5cm]  
Measurement of the Branching Fraction for
$B^0\to\aonepi$ with 535 Million $B\overline{B}$ Pairs
}

\affiliation{Budker Institute of Nuclear Physics, Novosibirsk}
\affiliation{Chiba University, Chiba}
\affiliation{University of Cincinnati, Cincinnati, Ohio 45221}
\affiliation{Department of Physics, Fu Jen Catholic University, Taipei}
\affiliation{Justus-Liebig-Universit\"at Gie\ss{}en, Gie\ss{}en}
\affiliation{The Graduate University for Advanced Studies, Hayama}
\affiliation{Gyeongsang National University, Chinju}
\affiliation{Hanyang University, Seoul}
\affiliation{University of Hawaii, Honolulu, Hawaii 96822}
\affiliation{High Energy Accelerator Research Organization (KEK), Tsukuba}
\affiliation{Hiroshima Institute of Technology, Hiroshima}
\affiliation{University of Illinois at Urbana-Champaign, Urbana, Illinois 61801}
\affiliation{Institute of High Energy Physics, Chinese Academy of Sciences, Beijing}
\affiliation{Institute of High Energy Physics, Vienna}
\affiliation{Institute of High Energy Physics, Protvino}
\affiliation{Institute for Theoretical and Experimental Physics, Moscow}
\affiliation{J. Stefan Institute, Ljubljana}
\affiliation{Kanagawa University, Yokohama}
\affiliation{Korea University, Seoul}
\affiliation{Kyoto University, Kyoto}
\affiliation{Kyungpook National University, Taegu}
\affiliation{Swiss Federal Institute of Technology of Lausanne, EPFL, Lausanne}
\affiliation{University of Ljubljana, Ljubljana}
\affiliation{University of Maribor, Maribor}
\affiliation{University of Melbourne, School of Physics, Victoria 3010}
\affiliation{Nagoya University, Nagoya}
\affiliation{Nara Women's University, Nara}
\affiliation{National Central University, Chung-li}
\affiliation{National United University, Miao Li}
\affiliation{Department of Physics, National Taiwan University, Taipei}
\affiliation{H. Niewodniczanski Institute of Nuclear Physics, Krakow}
\affiliation{Nippon Dental University, Niigata}
\affiliation{Niigata University, Niigata}
\affiliation{University of Nova Gorica, Nova Gorica}
\affiliation{Osaka City University, Osaka}
\affiliation{Osaka University, Osaka}
\affiliation{Panjab University, Chandigarh}
\affiliation{Peking University, Beijing}
\affiliation{University of Pittsburgh, Pittsburgh, Pennsylvania 15260}
\affiliation{Princeton University, Princeton, New Jersey 08544}
\affiliation{RIKEN BNL Research Center, Upton, New York 11973}
\affiliation{Saga University, Saga}
\affiliation{University of Science and Technology of China, Hefei}
\affiliation{Seoul National University, Seoul}
\affiliation{Shinshu University, Nagano}
\affiliation{Sungkyunkwan University, Suwon}
\affiliation{University of Sydney, Sydney, New South Wales}
\affiliation{Tata Institute of Fundamental Research, Mumbai}
\affiliation{Toho University, Funabashi}
\affiliation{Tohoku Gakuin University, Tagajo}
\affiliation{Tohoku University, Sendai}
\affiliation{Department of Physics, University of Tokyo, Tokyo}
\affiliation{Tokyo Institute of Technology, Tokyo}
\affiliation{Tokyo Metropolitan University, Tokyo}
\affiliation{Tokyo University of Agriculture and Technology, Tokyo}
\affiliation{Toyama National College of Maritime Technology, Toyama}
\affiliation{Virginia Polytechnic Institute and State University, Blacksburg, Virginia 24061}
\affiliation{Yonsei University, Seoul}
 \author{K.~Abe}\affiliation{High Energy Accelerator Research Organization (KEK), Tsukuba} 
 \author{I.~Adachi}\affiliation{High Energy Accelerator Research Organization (KEK), Tsukuba} 
 \author{H.~Aihara}\affiliation{Department of Physics, University of Tokyo, Tokyo} 
 \author{D.~Anipko}\affiliation{Budker Institute of Nuclear Physics, Novosibirsk} 
 \author{K.~Arinstein}\affiliation{Budker Institute of Nuclear Physics, Novosibirsk} 
 \author{T.~Aso}\affiliation{Toyama National College of Maritime Technology, Toyama} 
 \author{V.~Aulchenko}\affiliation{Budker Institute of Nuclear Physics, Novosibirsk} 
 \author{T.~Aushev}\affiliation{Swiss Federal Institute of Technology of Lausanne, EPFL, Lausanne}\affiliation{Institute for Theoretical and Experimental Physics, Moscow} 
 \author{T.~Aziz}\affiliation{Tata Institute of Fundamental Research, Mumbai} 
 \author{S.~Bahinipati}\affiliation{University of Cincinnati, Cincinnati, Ohio 45221} 
% ...
\collaboration{The Belle Collaboration}

\begin{abstract}
We present a measurement of the branching fraction
for the decay $B^0\to\aonepi$
with $\aonedecay$
using a data sample containing $535\times 10^6$ $B\overline{B}$
pairs collected with the Belle detector at the KEKB 
asymmetric-energy $e^+e^-$ collider operating at the $\Upsilon(4S)$
resonance.
We measure the branching fraction 
${\cal B}(B^0\to\aonepi){\cal B}(\aonedecay)
=(\aonepibr \pm \aonepibrstat \pm \aonepibrsyst )\times 10^{-6}$,
where the first and second errors are statistical and systematic,
respectively.
\end{abstract}


\maketitle

\tighten

{\renewcommand{\thefootnote}{\fnsymbol{footnote}}}
\setcounter{footnote}{0}

The decay $B^0\to\aonepi$~\cite{CC} proceeds through $b\to u$ transitions,
hence its time-dependent $CP$ violation is sensitive to
the Cabibbo-Kobayashi-Maskawa (CKM)~\cite{ckm} angle $\phi_2$~\cite{phi2}.
The first attempt to search for this decay mode was made
by the CLEO collaboration; an upper limit of $490\times 10^{-6}$
was obtained at the 90\% confidence level (C.L.)~\cite{cleo}.
Recently, the BaBar collaboration reported the first measurement
of the branching fraction
of $B^0\to\aonepi$ assuming a 50\% branching fraction for
$\aonedecay$ decay:
$(33.2\pm3.8\pm3.0)\times 10^{-6}$, where the first (second)
error is statistical (systematic) error~\cite{babar-br}.
The measured branching fraction is compatible with the prediction
in Ref.~\cite{br-prediction}.
BaBar also measured 
the time-dependent $CP$-violating parameters~\cite{babar-tcpv}.
The $CP$-violating parameters provide an effective $\phi_2$ value
that can be shifted from the CKM angle $\phi_2$ due to the
contribution from $b\to d$ ``penguin" decay amplitudes.
The value of $\phi_2$ can be extracted using the method
proposed in Ref.~\cite{aonepi-phi2}.

In this report, we present a measurement of the branching fraction 
of the decay $B^0\to\aonepi$ with $\aonedecay$
based on a data sample containing 
$N_{B\overline{B}}=(535\pm7) \times 10^6$ $B\overline{B}$ pairs.
It is known that the $\aone$ decays into
$\rho^0\pi^{\pm}$ via both $S$ and $D$-waves, 
and $\sigma(600)\pi^{\pm}$~\cite{pdg}.
In this analysis, we use a Monte Carlo (MC) simulation
with $S$-wave $\aone\to\rho^0\pi^{\pm}$ decays to
estimate the nominal detection efficiency.
We then assign a systematic error on the detection efficiency 
from the contributions of
$\aone$ decays into $D$-wave $\rho^0\pi^{\pm}$ and $\sigma(600)\pi^{\pm}$.
The non-resonant contributions from $B^0\to \rho^0\pi^+\pi^-$
and $\pi^+\pi^-\pi^+\pi^-$ decays are neglected in the nominal fit.
We estimate the possible contribution of these modes from a fit
to the three pion mass distribution
and assign a corresponding systematic error.

The data sample was collected with the Belle detector~\cite{belle-detector}
at the KEKB asymmetric-energy $e^+ e^-$ (3.5 on 8 GeV) collider~\cite{kekb}
operating at the $\Upsilon(4S)$ resonance.
The Belle detector
is a large-solid-angle magnetic spectrometer 
that consists of a silicon vertex detector (SVD),
a 50-layer central drift chamber (CDC),
an array of aerogel threshold Cherenkov counters (ACC),
a barrel-like arrangement of time-of-flight scintillation 
counters (TOF),
and an electromagnetic calorimeter (ECL)
comprised of CsI(Tl) crystals located inside a superconducting
solenoid coil that provides a 1.5~T magnetic field.
An iron flux-return located outside of the coil is 
instrumented to detect $K^0_L$ mesons and to identify muons (KLM).
A sample containing $152 \times 10^6$ $B\overline{B}$ pairs 
was collected with a 2.0~cm radius beampipe and a 3-layer
silicon vertex detector, while
a sample with $383 \times 10^6$ $B\overline{B}$ pairs 
was collected with a 1.5~cm radius beampipe, 
a 4-layer silicon detector, and
a small-cell inner drift chamber.

We reconstruct $B^0\to\aonepi$ candidates from 
combinations of four charged tracks
originating from the beam interaction region.
The tracks are required to be consistent with a pion hypothesis
based on the particle identification (PID) information
from the ACC and the $dE/dx$ measurements in the CDC.
Positively identified electrons are rejected.

An $\aone$ candidate is reconstructed from three charged pions 
$\pi^{\pm}\pi^{\pm}\pi^{\mp}$ with invariant mass in the range
from 0.8~GeV/c$^2$ to 1.8~GeV/c$^2$.
We require at least one pair of oppositely charged pion candidates
satisfy the condition 
$0.55$~GeV/c$^2<m_{\pi^+\pi^-}<1.15$~GeV/c$^2$,
where $m_{\pi^+\pi^-}$ is the invariant mass of the pair.

To construct $B^0$ candidates we combine $\aone$ candidates with
a bachelor pion having momentum in the range
$2.2$~GeV/c $<p_{\rm bach}<2.7$~GeV/c in the
$\Upsilon(4S)$ center-of-mass system (CMS).
The $B^0$ candidates are identified using
the energy difference $\de \equiv E_B^* - E_{\rm beam}^*$
and the beam energy constrained mass
$\mbc\equiv\sqrt{(E_{\rm beam}^*)^2-(p_B^*)^2}$,
where $E_{\rm beam}^*$ is the CMS beam-energy,
and $E_B^*$ and $p_B^*$ are the CMS energy and momentum of
the $B^0$ candidate.
We select the $B^0$ candidates in the region
$5.20$~GeV/c$^2$ $<\mbc<5.30$~GeV/c$^2$ and
$|\de|<0.12$~GeV.

We explicitly eliminate three charm $B$ decay modes that peak
in the $\mbc$ signal region:
$B^0\to D^{-}\pi^+$ with $D^{-}\to K^+ \pi^-\pi^-$,
$B^0\to D^{*-}\pi^+$ with $D^{*-}\to \overline{D}{}^0\pi^-$ 
and $\overline{D}{}^0\to K^+\pi^-$,
and $B^0\to J/\psi K^{*0}$ with $J/\psi\to \mu^+\mu^-$ and
$K^{*0}\to K^+\pi^-$.
We exclude candidates in the mass ranges:
$1.86$~GeV/c$^2$ $<M_{D^-}<1.88$~GeV/c$^2$,
$2.00$~GeV/c$^2$ $<M_{D^{*-}}<2.05$~GeV/c$^2$,
and $3.05$~GeV/c$^2$ $<M_{J/\psi}<3.15$~GeV/c$^2$.

The dominant background for this analysis comes from 
continuum events of the type $e^+e^-\to q\overline{q}$ $(q=u,d,s,c)$.
To suppress this background, we construct likelihood functions,
$L_{S}$ and $L_{BG}$,
for the signal $B^0\to\aonepi$ decay 
and the continuum background, respectively.
The likelihood functions are formed as a product
of probability density functions (PDF) of
a Fisher discriminant formed from event shape
variables~\cite{sfw} and the angle of the $B$ candidate
flight direction in the CMS with respect to the beam axis.
We also make use of the variable $r$ provided by 
a flavor tagging algorithm~\cite{flavor-tag}
that identifies the flavor of the accompanying $B^0$ in the 
$\Upsilon(4S)\to B^0\overline{B}{}^0$ decay.
The parameter $r$ ranges from $r=0$ for no flavor discrimination
to $r=1$ for unambiguous flavor assignment.
The data sample is divided into six $r$ intervals.
Since the separation of the continuum background from the signal 
depends on $r$,
we determine the requirement on the likelihood ratio 
${\cal R} = L_{S}/(L_{S}+L_{BG})$ for each $r$ bin
by optimizing the expected sensitivity
using signal MC events 
and events in the sideband region $\mbc<5.26$~GeV/c$^2$.

We find that on average 2.9 $B^0\to\aonepi$ decay candidates 
are included for each event selected from the data. 
The $B^0$ candidate having the largest ${\cal R}$ value 
is chosen in the case of multiple candidates in an event.
After the best candidate selection,
we find that 19.9\% of the signal events in the MC simulation
are incorrectly reconstructed self-cross feed (SCF) events; 
for those events at least one
charged pion track is replaced with 
one from the accompanying $B$ meson decay.
In 99.6\% of the SCF events
one or more tracks from the $\aonedecay$
has been replaced;
in the remaining 0.4\% a bachelor pion was replaced.
For the correctly reconstructed signal events, 
there is a possibility that the bachelor pion could be swapped with
a pion from the $\aonedecay$ decay.
We find that the probability of obtaining the wrong combination 
is negligibly small;
the bachelor pion is unambiguously determined.

We extract the signal yield by using $\de$, $\mbc$ 
and $\cos\theta$, where $\theta$ is defined as the angle
between the normal to the $\aone$ decay plane and the bachelor
pion direction in the $\aone$ rest frame.
This angle is employed to discriminate between the signal 
and $B^0\to\atwo \pi^{\mp}$ events, 
which have the same distributions in $\de$ and $\mbc$
as the signal events.

For the correctly-reconstructed signal and $B^0\to\atwopi$ components,
we use a sum of two bifurcated Gaussians and Gaussian functions
to describe the $\de$ and $\mbc$ shapes, respectively.
The possible differences between data and MC in
the shapes of $\de$ and $\mbc$ distributions are
taken into account
using a $B^0\to D^-\pi^+$, $D^-\to K^+\pi^-\pi^-$
decay control sample.
The $\cos\theta$ distributions are polynomial functions obtained from 
the MC samples.
The $\de$-$\mbc$ and $\cos\theta$ PDFs for SCF events
are modeled by a smoothed two-dimensional histogram
and a polynomial, respectively.

The PDFs for charm $b\to c$ and charmless $b\to u$ backgrounds
are obtained from a large MC sample and 
modeled as two-dimensional smoothed histograms for $\de$-$\mbc$
and polynomial functions for $\cos\theta$.
The charmless $b\to u$ decay background contains a peaking background
that has the same shape in $\de$-$\mbc$ 
as that of the signal.
We estimate the peaking background yield using
the MC simulation, and find that the dominant $B$ decay modes
are $B^0\to K^{*+}\pi^-$, $K_0^{*+}\pi^-$ and $\rho^0\rho^0$.
For these modes 32 events are expected.
The uncertainty in this background component 
is included in the systematic error.

We model the continuum background event shapes
as a second-order polynomial and an ARGUS function~\cite{argus} 
for $\de$ and $\mbc$, respectively.
The parameters of the functions are determined from the fit.
We model the $\cos\theta$ distribution as a polynomial function.
The parameters of the functions are obtained using events in
the $\mbc$ sideband region defined as $\mbc < 5.26$~GeV/c$^2$.
The parameters are fixed in the fit.

\begin{figure}[htb]
\begin{center}
\includegraphics[width=0.35\textwidth]{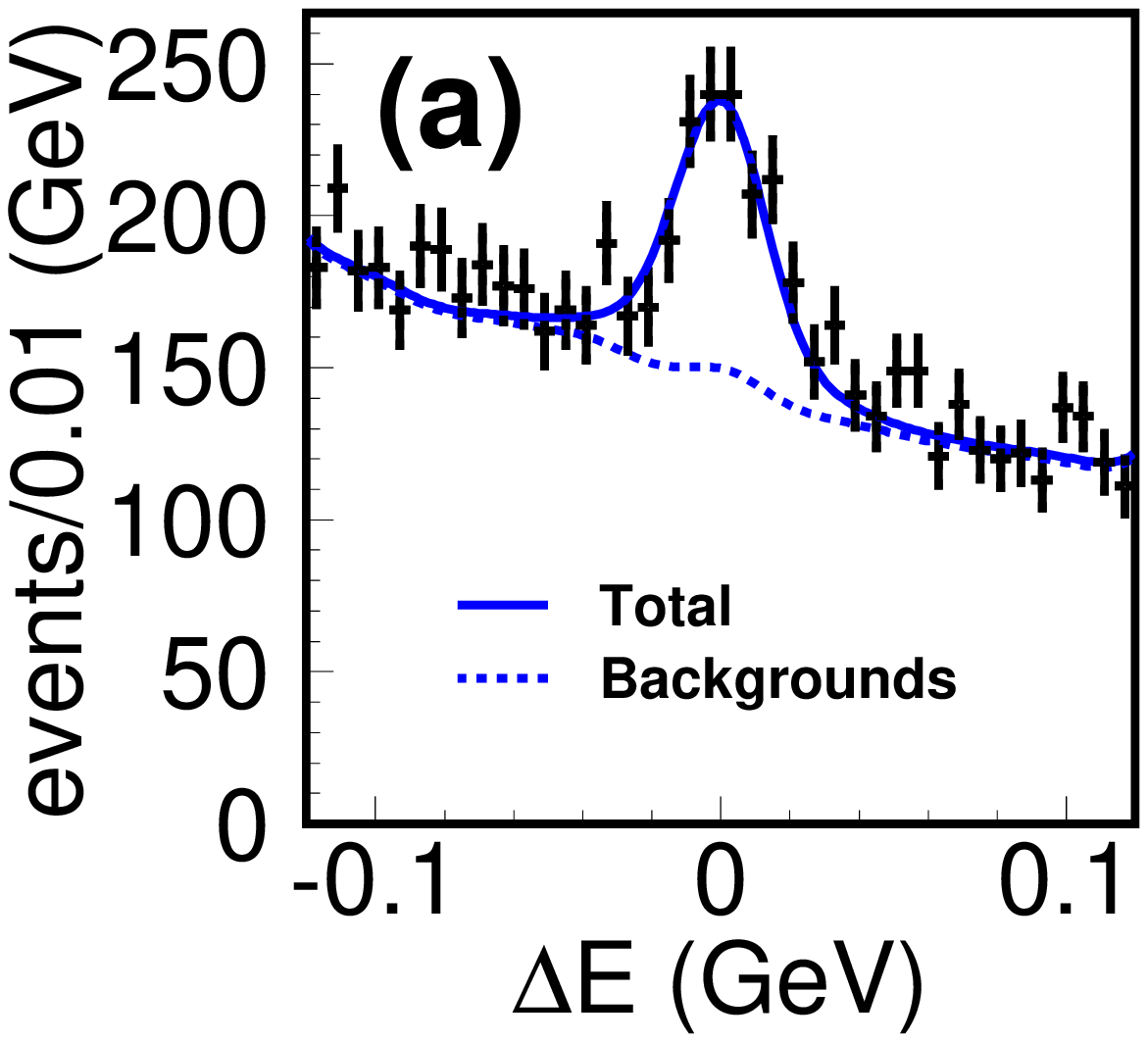}
\includegraphics[width=0.37\textwidth]{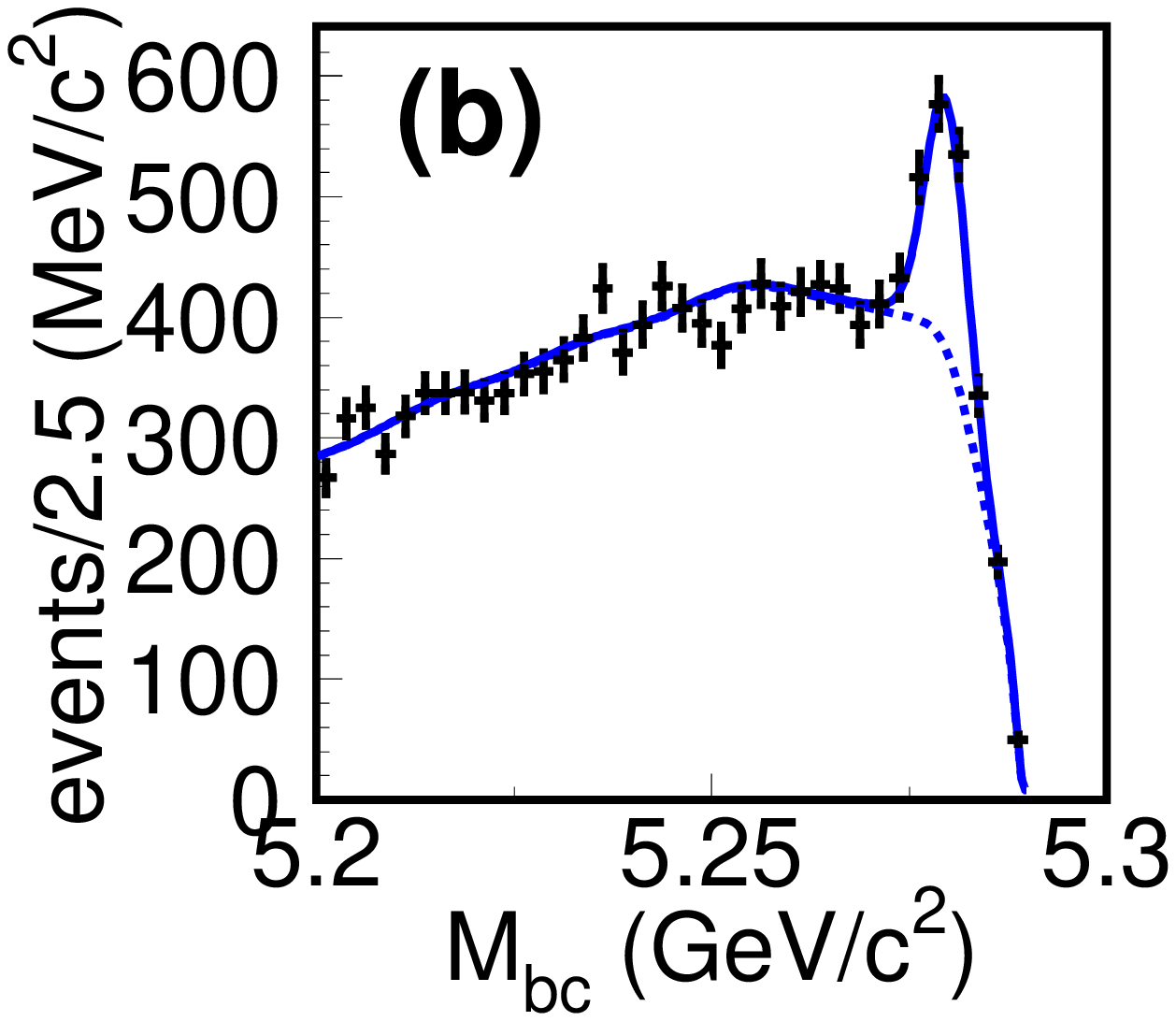}
\includegraphics[width=0.35\textwidth]{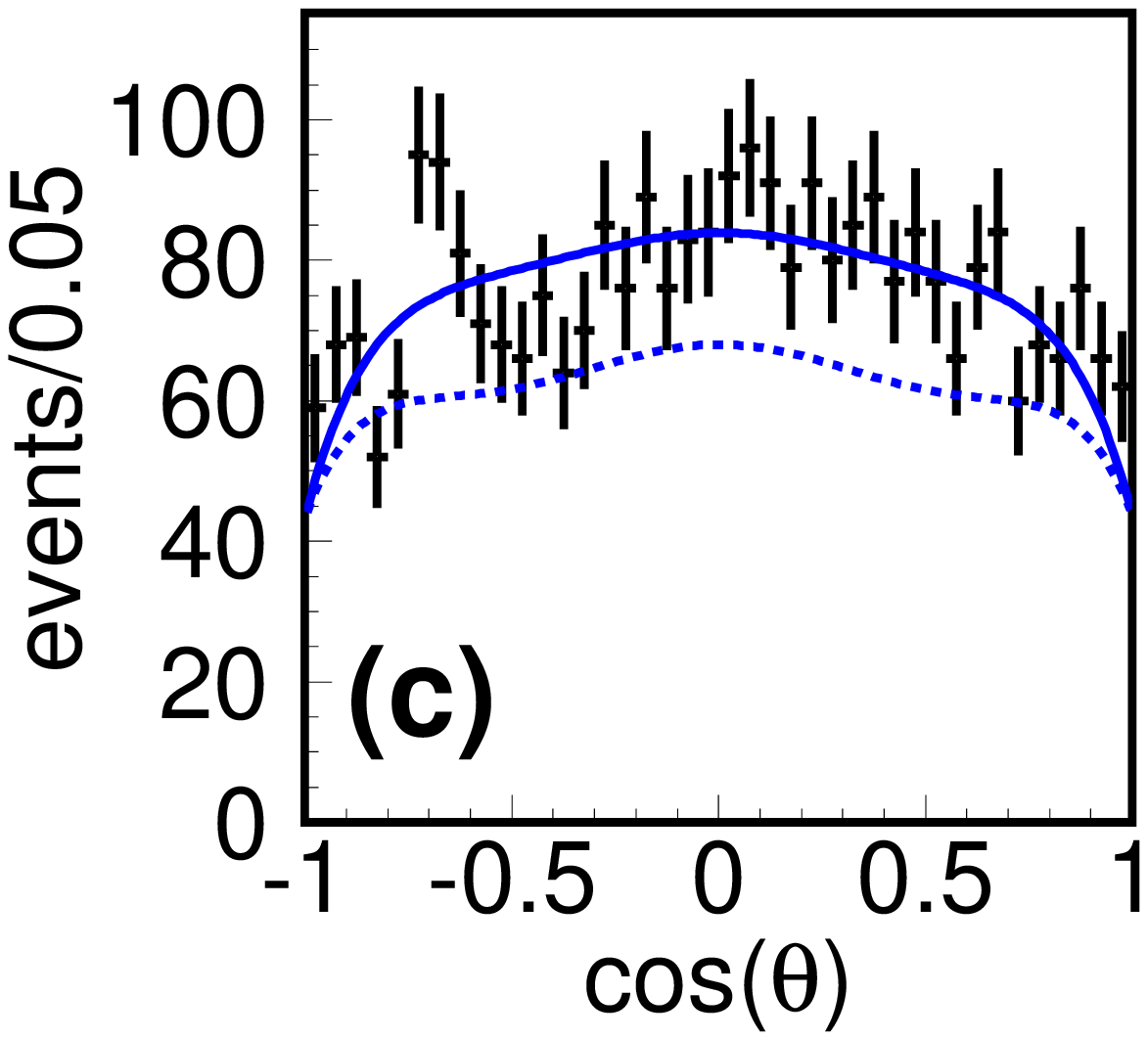}
\caption{(a) projection in $\de$ for events with $\mbc>5.27$~GeV/c$^2$,
(b) projection in $\mbc$ for events with $|\de|<0.05$~GeV,
and (c) projection in $\cos\theta$ for events 
with $\mbc>5.27$~GeV/c$^2$ and $|\de|<0.05$~GeV.
The points with error bars are the data.
The solid (dashed) curve shows the fit results,
which includes the sum of all the components (backgrounds).
}
\label{fig:fit-result}
\end{center}
\end{figure}

We perform a 3D ($\de$-$\mbc$-$\cos\theta$) unbinned extended 
maximum likelihood fit
to 31725 candidates.
The likelihood value is 
\begin{equation}
L=\exp(-\sum_j n_j)\prod_i[\sum_j n_j P_j(\de_i, M_{{\rm bc},i}, \cos\theta_i)],
\end{equation}
where $i$ runs over all the $B^0\to\aonepi$ candidates,
$j$ indicates one of the following event categories:
correctly reconstructed signal, SCF signal, 
the correctly reconstructed $B^0\to\atwopi$,  SCF $B^0\to\atwopi$,
$b\to c$, $b\to u$, and continuum backgrounds.
$n_j$ is the yield of each category $j$,
and $P_j(\de, \mbc, \cos\theta)$ is the PDF
of the category $j$ as a function of $\de$, $\mbc$ and $\cos\theta$.
The yields of the correctly reconstructed and SCF signal events
are parameterized as $(1-f_{\rm SCF})n_{a_1\pi}$
and $f_{\rm SCF}n_{a_1\pi}$, respectively,
where $f_{\rm SCF}=0.199$ is the SCF fraction determined 
using the signal MC simulation, and $n_{a_1\pi}$ is the 
signal yield.
We fix the $f_{\rm SCF}$ value in the fit.
The same parameterization is applied to the $B^0\to\atwopi$ decay mode,
where the SCF fraction (0.158) is also determined from the MC.
In the 3D $\de$-$\mbc$-$\cos\theta$ fit,
we vary the yields $n_j$ and parameters that determine
the continuum background shape in $\de$ and $\mbc$.
The fit 
yields $n_{a_1\pi} = 654\pm70$ signal events
and $n_{a_2\pi} = 47\pm50$ $B^0\to\atwopi$ events.
Figure~\ref{fig:fit-result} shows 
the $\de$, $\mbc$ and $\cos\theta$ distributions with
projections of the fit results.

We calculate the product of the branching fractions using
\begin{equation}
{\cal B}(B^0\to\aonepi){\cal B}(\aonedecay) 
= \frac{n_{a_1\pi}}
       {N_{B\overline{B}}\cdot\varepsilon_{\rm det}\cdot\varepsilon_{\rm PID}},
\end{equation}
where $\varepsilon_{\rm det}= 9.06$\% is the detection efficiency
estimated using the signal MC simulation assuming 
$\aone\to\rho^0\pi^{\pm}$ decay via an $S$-wave.
The correction factor $\varepsilon_{\rm PID}=0.90$ 
takes into account the PID selection efficiency difference
between the real data and the MC simulation. It is obtained from
a large sample of $D^{*+}\to D^{0}\pi^{+}$, $D^{0}\to K^-\pi^+$ decays
in data.
We measure the product branching fraction to be
$(\aonepibr\pm\aonepibrstat)\times 10^{-6}$, where the error is statistical.

To validate our results,
we measure the branching fraction of the decay
$B^0\to D^-\pi^+$, $D^-\to K^+\pi^-\pi^-$ as a control sample.
We obtain 
${\cal B}(B^0\to D^-\pi^+) = (2.42\pm0.05)\times 10^{-4}$, where the error
is statistical only.
The value is consistent with the World Average (W.A.) value
of $(2.54\pm0.28)\times 10^{-4}$~\cite{pdg}.
To estimate the fit bias, we perform fits to an ensemble
of toy-MC pseudo-experiments and MC samples.
The signal-to-background fractions used are
obtained from the fit to the data.
No bias is found in the fits to pseudo-experiment samples.
We find a small difference ($3.8\pm5.3$\%) for the fit to the MC samples,
which is also consistent with zero.
We assign a systematic uncertainty of 
$\pm3.8$\% to allow for this.

\begin{figure}[htb]
\begin{center}
\includegraphics[width=0.6\textwidth]{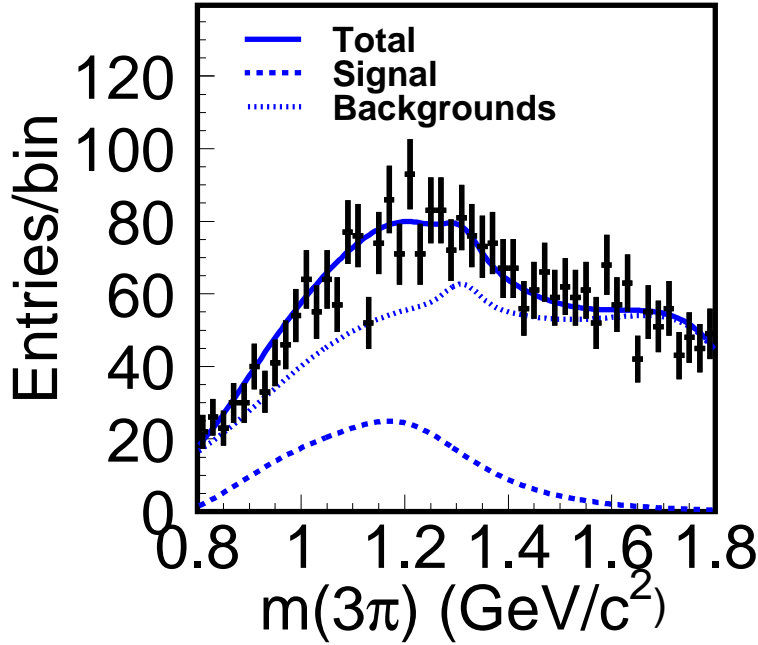}
\caption{
Three pion mass distribution with the fit result.
The dashed and dotted curves show the signal and the sum of backgrounds,
respectively.
The solid curve indicates the sum of all the components.
The points with error bars are the data.
}
\label{fig:a1mass}
\end{center}
\end{figure}

We determine the $\aone$ mass ($M_{a_1}$) and width ($\Gamma_{a_1}$)
from an unbinned maximum likelihood fit to
the $\aonedecay$ three pion mass distribution of the 2915 
candidates in the signal box defined as $|\de|<0.05$~GeV and
$\mbc>5.27$~GeV/c$^2$.
We employ a relativistic Breit-Wigner (BW) function
to model the $\aone$ mass distribution~\cite{bw}, which is multiplied
by the mass dependent efficiency to form a PDF 
for the correctly reconstructed signal events.
We use a MC-determined PDF for the signal SCF events
with the mass-dependent weight determined by the BW function.
For the correctly reconstructed $B^0\to\atwopi$ events, 
we use a sum of three Gaussians.
Since the $B^0\to\atwopi$ SCF contribution 
is very small ($<0.1$\%) in the signal box,
we neglect it.
The PDFs of the charm and charmless $B$ backgrounds are determined using 
the MC samples.
The continuum PDF is obtained from the $\mbc$ sideband events
by subtracting the charm $B$ decay contamination.
In the fit, only two parameters, $M_{a_1}$ and $\Gamma_{a_1}$, are floated.
The fractions of the signal to the backgrounds are fixed 
to the value determined from the 3D $\de$-$\mbc$-$\cos\theta$ fit.
The fit yields $M_{a_1}=1233^{+84}_{-47}$~MeV/c$^2$
and $\Gamma_{a_1} = 594^{+446}_{-152}$~MeV/c$^2$,
where the errors are statistical only.
The measured $\aone$ mass and width are comparable 
with the W.A. values~\cite{pdg}.
Figure~\ref{fig:a1mass} shows the three pion mass distribution
along with the fit results.

To estimate possible contributions from 
non-resonant $B^0$ decays into 
$\rho^0\pi^+\pi^-$ and $\pi^+\pi^-\pi^+\pi^-$,
we make use of the three pion mass distribution,
since the 3D $\de$-$\mbc$-$\cos\theta$ fit cannot
distinguish between the signal and non-resonant components.
We include PDFs for the non-resonant decay modes in the fitting
likelihood function.
The PDF of the non-resonant components is determined from large MC samples
generated assuming three- and four-body phase space distributions;
we use a single PDF for both modes since
the shapes of these distributions are nearly the same.
We divide the correctly reconstructed signal fraction in the signal box, 
$f_{\rm sig}=17.2$\%, 
into $\aonepi$ and non-resonant contributions;
the $\aonepi$ and non-resonant fractions in the fit are parameterized as
$f_{\rm sig}(1-f_{\rm nr})$ and $f_{\rm sig}f_{\rm nr}$, respectively,
where $f_{\rm nr}$ is the non-resonant fraction. 
The fraction $f_{\rm nr}$, $M_{a_1}$ and $\Gamma_{a_1}$
are allowed to float in the fit.
We obtain $f_{\rm nr} = 0.11\pm0.07$, and $M_{a_1}$ and $\Gamma_{a_1}$
values consistent with the results from 
the nominal fit where $f_{\rm nr} = 0$.

The dominant systematic error of the branching fraction
originates from the uncertainty in the non-resonant contributions,
which corresponds to a $\pm 11$\% systematic error.
By varying the PDF parameters,
we estimate the PDF systematic uncertainty of $\pm6.9$\%. 
We assign a $\pm 4.8$\% tracking reconstruction efficiency uncertainty. 
The systematic error on the detection efficiency mainly originates from
the uncertainties in the $\aone$ decays into 
$\rho^0\pi^{\pm}$ via a $D$-wave.
We assign a $\pm4.7$\% systematic error due to these uncertainties.
We vary the branching fractions of the peaking backgrounds 
by one standard deviation,
and add the differences from the nominal fit result in quadrature
to determine systematic uncertainties of $\pm1.7$\%.
Other sources of systematic error are
the uncertainties in PID selection efficiency ($\pm 1.3$\%),
fit bias ($\pm 3.8$\%)
and number of $B\overline{B}$ pairs ($\pm 1.3$\%).
We add each contribution in quadrature to obtain
the total systematic error of $\pm 15.3$\%.

In summary, we measure the branching fraction product 
of $B^0\to\aonepi$ and $\aonedecay$
using a data sample containing 
$535\times 10^6$ $B\overline{B}$ pairs:
${\cal B}(B^0\to\aonepi){\cal B}(\aonedecay)
= (\aonepibr\pm\aonepibrstat\pm \aonepibrsyst)\times 10^{-6}$, 
where the first and second errors are statistical and systematic, 
respectively.
If we assume a 50\% branching fraction for $\aonedecay$ decays,
${\cal B}(B^0\to\aonepi) = (29.8\pm3.2\pm 4.6)\times 10^{-6}$.
Our result is consistent with other measurements~\cite{babar-br}.

We thank the KEKB group for the excellent operation of the
accelerator, the KEK cryogenics group for the efficient
operation of the solenoid, and the KEK computer group and
the National Institute of Informatics for valuable computing
and Super-SINET network support. We acknowledge support from
the Ministry of Education, Culture, Sports, Science, and
Technology of Japan and the Japan Society for the Promotion
of Science; the Australian Research Council and the
Australian Department of Education, Science and Training;
the National Science Foundation of China and the Knowledge
Innovation Program of the Chinese Academy of Sciences under
contract No.~10575109 and IHEP-U-503; the Department of
Science and Technology of India; 
the BK21 program of the Ministry of Education of Korea, 
the CHEP SRC program and Basic Research program 
(grant No.~R01-2005-000-10089-0) of the Korea Science and
Engineering Foundation, and the Pure Basic Research Group 
program of the Korea Research Foundation; 
the Polish State Committee for Scientific Research; 
the Ministry of Education and Science of the Russian
Federation and the Russian Federal Agency for Atomic Energy;
the Slovenian Research Agency;  the Swiss
National Science Foundation; the National Science Council
and the Ministry of Education of Taiwan; and the U.S.\
Department of Energy.


\begin{thebibliography}{99}
\bibitem{CC}
Throughout this paper, 
the inclusion of the charge conjugate mode decay is implied
unless otherwise stated.
\bibitem{phi2} Another naming convention, 
$\alpha$ ($= \phi_2$) is also used in the literature.
\end{thebibliography}
\end{document}